\documentclass{article}

\usepackage{amsmath,amssymb,amsthm,bbm,color,hyperref,nccmath}
\usepackage[margin=1in]{geometry}
\usepackage{caption}
\usepackage{graphicx}
\usepackage{subcaption}
\usepackage{afterpage}
\usepackage[hang,flushmargin]{footmisc}
\usepackage{float}
\usepackage{enumitem,kantlipsum}
\usepackage{multirow}
\usepackage{algorithm}
\usepackage{algpseudocode}
\usepackage{booktabs}

\usepackage[round]{natbib}

\newtheorem{theorem}{Theorem}

\newtheorem{corollary}{Corollary}

\newtheorem{assumption}{Assumption}

\newcommand{\EE}{\mathbb{E}}

\newcommand{\VV}{\mathbb{V}}
\newcommand{\PP}{\mathbb{P}}

\newcommand{\One}[1]{{\mathbbm{1}}\left\{{#1}\right\}}

\allowdisplaybreaks

\title{A Multiplicative Instrumental Variable Model\\ for Data Missing Not-at-Random}
\author{Yunshu Zhang, Chan Park, Jiewen Liu, Yonghoon Lee, Mengxin Yu, \\
James M. Robins, Eric J. Tchetgen Tchetgen}

\begin{document}

\maketitle

\begin{abstract}
Instrumental variable (IV) methods offer a valuable approach to account for outcome data missing not-at-random.  A valid missing data instrument is a measured factor which (i) predicts the nonresponse process and (ii) is independent of the outcome in the underlying population. For point identification, all existing IV methods for missing data including the celebrated Heckman selection model, a priori restrict the extent of selection bias on the outcome scale, therefore potentially understating uncertainty due to missing data. In this work, we introduce an  IV framework which allows the degree of selection bias on the outcome scale to remain completely unrestricted. The new approach instead relies for identification on (iii) a key multiplicative selection model,  which posits that the instrument and any hidden common correlate of selection and the outcome, do not interact on the multiplicative scale. Interestingly, we establish that any regular statistical functional of the missing outcome is nonparametrically identified under (i)-(iii)  via a \emph{single-arm Wald ratio estimand} reminiscent of the standard Wald ratio estimand in causal inference.  For estimation and inference, we characterize the influence function for any functional defined on a nonparametric model for the observed data, which we leverage to develop semiparametric multiply robust IV estimators. Several extensions of the methods are also considered, including the important practical setting of polytomous and continuous instruments. Simulation studies illustrate the favorable finite sample performance of proposed methods, which we further showcase in an HIV study nested within a household health survey study we conducted in Mochudi, Botswana, in which interviewer characteristics are used as instruments to correct for selection bias due to dependent nonresponse in the HIV component of the survey study.
\end{abstract}

\section{Introduction}

Missing data is a common and critical issue that impacts most empirical studies conducted in the health and social sciences. Failure to address selection bias induced by missing data can lead to biased and inconsistent results. \citet{rubin1976inference} was an early proponent for formal statistical methods for addressing missing data, including introducing language for classifying missing data mechanisms into three broad categories: missing completely at random (MCAR), missing at random (MAR), and missing not at random (MNAR). 
In a setting where the sampled data consists of an outcome variable of primary interest which is potentially missing, and a set of fully observed baseline covariates, MCAR assumes that the missingness mechanism is completely independent of both the outcome and covariates, in which case a standard complete-case approach to handle missing data, which restricts analyses, e.g. a regression model of the outcome on covariates, to the subset of observations with fully observed data is guaranteed to be unbiased, although potentially inefficient. In contrast, under MAR, one assumes that conditional on fully observed covariates, the missingness mechanism is independent of the outcome variable. Under this assumption, most well-defined statistical functionals of the the full data are identifiable from the observed data using methods such as complete-case inverse probability weighting, direct maximum likelihood inference, multiple imputation and augmented inverse probability weighting 
\citep{seaman2012combining, rubin2018multiple, van2011targeted, robins1994estimation, tsiatis2006semiparametric}
, all of which are viable options for obtaining valid inference, while formally accounting for selection bias. Lastly, missing data are said to be MNAR, if say, an unmeasured factor induces missingness and is also associated with the outcome, rendering the two dependent. In such setting, identification can be significantly more challenging, and the aforementioned methods designed for MCAR or MAR are generally invalid due to the latent dependence between the missingness mechanism and the unobserved outcome. Unfortunately, whether missing data mechanisms are MAR or MNAR is untestable from the observed data alone without a different strong assumption. As a result, there has been growing interest over the recent years in the development of statistical methods that can accommodate MNAR mechanisms.

To identify a target population functional or parameter of interest under MNAR, some existing statistical methods have relied on strong parametric model restrictions, which can be sensitive to even slight specification error \citep{diggle1994informative, wu1988estimation, roy2003modeling, miao2016identifiability}.
Alternatively, sensitivity analysis has been proposed as a framework to assess the potential impact of MNAR mechanisms on inference in nonparametric and semiparametric models  \citep{rotnitzky1997analysis,robins2000sensitivity}. 
A different strand of work has considered nonparametric identification and semiparametric inference leveraging a so-called shadow variable, apriori known to be independent of the missingness mechanism conditional on the potential unobserved outcome variable and measured covariates \citep{miao2024identification, kott2014calibration, miao2016varieties, li2023non}. 
While, partly inspired by methods originally proposed by \citet{heckman1979sample,  heckman1997instrumental} 
and further developed by \cite{das2003nonparametric, newey1990semiparametricb}
,  \cite{tchetgen2017general} and \cite{sun2018semiparametric} considered an instrumental variable (IV) approach to account for selection bias for an outcome missing not at random. An IV in a missing outcome data setting is a fully observed variable, which is apriori known to be independent of the outcome in the underlying population, and is a strong correlate of the missingness process, conditional on observed covariates. Though Heckman's selection model relies on crucial parametric models specification to identify a population outcome regression function in view, \cite{sun2018semiparametric} established necessary and sufficient conditions for identifying the full data joint distribution within this IV framework; and  \cite{tchetgen2017general} considered a sufficient condition for IV identification of a regression function under a homogoneity condition which restricts the degree of selection bias on the scale of the outcome model.


In this paper, we propose a new multiplicative instrumental variable  (MIV) model for selection bias due to outcome missing not at random, which provides identification of any full data functional of the outcome, while avoiding the homogeneity restrictions of previous methods on the outcome model scale.  Our key identifying assumption is that the IV and any unobserved common correlate of both the missingness mechanism and the outcome, impact the selection probability via a product form, reflecting independent mechanisms of actions. 
For example, in a study estimating HIV seroprevalence among adults in Mochudi, Botswana, data were collected through a household survey \citep{novitsky2015phylodynamic}. Unfortunately, a substantial fraction of enumerated households had one or more participants who did not complete the HIV testing component of the survey visit. The concern is that, failure to test for HIV may be associated with unobserved health-seeking behaviors and other risk factors for HIV infection, in which case missing data on HIV status is most likely not at random. In this setting, because interviewers are assigned at random, an \emph{interviewer's characteristics}, such as years of experience as a survey field worker, years of education and other personal characteristics are strong predictors of whether or not an interviewee agrees to test for HIV, that are exogenous and therefore independent of an \emph{interviewee's unmeasured health-seeking behavior}; therefore interviewer characteristics are strong candidate instruments.  
Furthermore, it is unlikely that the mechanism by which an interviewer's characteristics might lead to nonresponse interacts with one by which an interviewee's individual risk factors for HIV infection might lead to nonresponse. Such \emph{independent mechanisms of action} is faithfully represented by the proposed multiplicative selection instrumental variable model. 

Our main contributions beyond nonparametric identification of any given smooth functional of the outcome distribution in the target population, including the outcome mean, include the characterization of the influence function for the functional of interest under a semiparametric full data model. In addition, we propose a multiply robust estimator for the population outcome mean functional with high-order asymptotic bias, and as a result, of any functional that can be expressed as the solution to a moment equation, under the proposed multiplicative selection IV model. 
Our work is closely related to and builds upon recent advances by \citet{liu2025multiplicativeinstrumentalvariablemodel, lee2025inferencenonlinearcounterfactualfunctionals}, where an analogous MIV model is proposed for making inferences about a causal effect. Crucially, we extend their results from causal inference to missing data problems, whereby, we generalize their MIV framework from the case of a single binary IV, to more general IV settings, where the instrument may be polytomous, continuous, or multivariate. As we demonstrate, these generalizations are not trivial extensions of prior results and require new technical developments, particularly with respect to the semiparametric estimation and efficiency theory, therefore providing several new results which are of independent interest both for missing data and causal IV settings.  

The rest of the paper is organized as follows. Section \ref{sec:miv} formally introduces notation for the paper and defines the  multiplicative selection model.   
Section \ref{sec:idandest} presents the identification and estimation results for the population mean under missingness, starting with the binary IV case. This section largely restates the results of \cite{liu2025multiplicativeinstrumentalvariablemodel} and \cite{lee2025inferencenonlinearcounterfactualfunctionals}, reformulated in missing-data notation. Section \ref{sec:generalIV} contains our main contributions, extending the identification and estimation results to the setting of general IVs. We derive the influence function  of a smooth functional of the outcome, and we establish the asymptotic properties of the corresponding semiparametric estimator.
Section \ref{sec:sim} demonstrates the performance of our proposed estimators through simulation studies. Section \ref{sec:realdat} applies our method to the HIV seroprevalence study in Mochudi, Botswana. Section \ref{sec:discussion} concludes with a discussion of potential future extensions.

\section{Multiplicative Selection Model}
\label{sec:miv}
\subsection{Notations and Assumptions}

Suppose one observes n independent and identically distributed (i.i.d) samples on $O=(X,RY,R,Z)$, such that $Y$ is not observed if $R=0$. Let  $X$ denote observed covariates, $Y$ is the outcome of interest, and $Z$ is the IV. $Y$ can be either continuous, binary, or polytomous, and $X$ is a vector with possibly all three types of variables.
Suppose that one aims to estimate a functional $\psi_{0}=\psi(F)$ of the full data distribution $F$ of $(X,Y)$, that uniquely solves the full data moment equation 
\[
0=\EE\left[ h\left( Y;\psi _{0}\right) \right] =\PP\left( R=1\right) \EE\left[
h\left( Y;\psi _{0}\right) |R=1\right] +\PP\left( R=0\right) \EE\left[
h\left( Y;\psi _{0}\right) |R=0\right], 
\]%
where $h$ is a pre-specified function and $\psi_{0}$ is the true value of some parameter of interest. The following are two standard examples.

\begin{figure}[htbp]
  \centering
\includegraphics[width=0.3\linewidth]{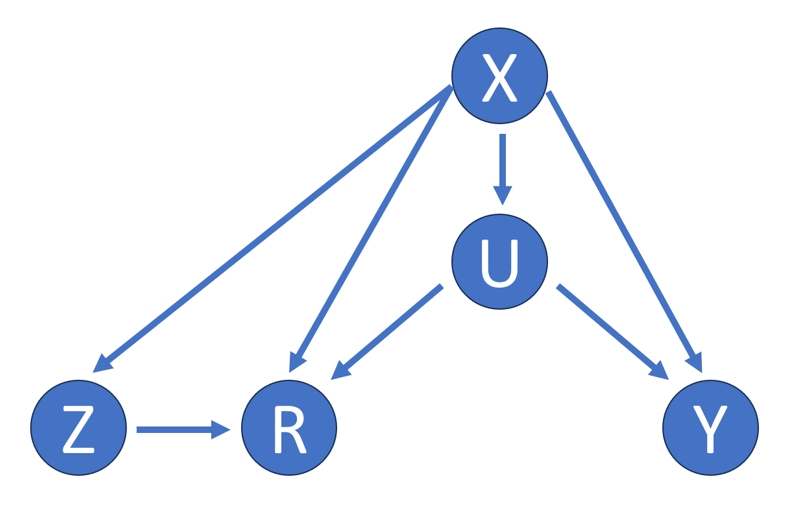}
    \caption{DAG of the multiplicative selection model with unmeasured confounders \( U \) included.\label{fig:dag1}}
\end{figure}

\begin{itemize}
    \item If $h\left( Y;\psi _{0}\right) = Y - \psi _{0}$, then $\psi _{0} = \EE\left[Y \right]$ is the population mean;
    \item If $h\left( Y;\psi _{0}\right) =  \One{Y \geq \psi _{0}} - q$, where $\One{\cdot}$ is the indicator function, then, supposing that $Y$ is continuous $\psi _{0} = F_{Y}^{-1}(q)$ is the $q$-th quantile of the population.
\end{itemize}
Suppose further that there exist latent factors $U$, which induce dependence between $Y$ and the missingness indicator $R$, such that the missingness mechanism is as a result MNAR.
Because $Y$ is observed for subjects with $R = 1$, the key challenge is to estimate the unobservable functional $\beta :=\EE\left[
h\left( Y;\psi _{0}\right) |R=0\right]$. 
To identify $\beta$, we assume the multiplicative IV model 
with dependence structure in the population illustrated in the following the Directed Acyclic Graph (DAG) in Figure \ref{fig:dag1}, which encodes the following missing data IV core conditions.

\begin{assumption}[weak ignorability]\label{assump:weakign}
    $Y$ independent of $Z,R$ given $(U,X)$.  
\end{assumption}

\begin{assumption}[conditional independence]\label{assump:condind}
    $U$ independent of $Z$ given $X$.
\end{assumption}

\begin{assumption}[multiplicative selection model and instrumental relevance]\label{assump:multi} 
$\\$$\PP\left( {R = 0|Z,U,X} \right) = \exp \left\{ {{\alpha _z}\left( {Z,X} \right) + {\alpha _u}\left( {U,X} \right)} \right\}$,
    where $\alpha_z(\cdot,\cdot)$ and $\alpha_u(\cdot,\cdot)$ are arbitrary functions satisfying ${\alpha _z}\left( {z,x} \right) \ne {\alpha _z}\left( {z',x} \right)$ for all $x$ and $z \neq z'$ where the natural constraint $\PP(R=0|Z,U,X) \in [0,1)$ holds almost surely.
\end{assumption}
Assumption \ref{assump:weakign} is quite a mild condition as $U$ is not required to be observed. Assumption \ref{assump:condind} indicates that the source of variation induced by the instrument must essentially be exogenous relative to the outcome process, a critical design consideration in selecting an appropriate instrument. 
Assumption \ref{assump:multi} formalizes the multiplicative selection model condition and IV relevance. 
An immediate implication of \ref{assump:multi}  is that $\tilde \pi \left( {z,u,x} \right) \ne \tilde \pi \left( {z',u,x} \right)$ for all $u, x$ and $z \ne z'$, 
    where $\tilde \pi \left( {Z,U,X} \right) = \PP\left( {R = 1|Z,U,X} \right)$; a standard IV relevance condition.

      

We use the following notations throughout the paper:
\begin{align*}
    &\pi_z \left( {X} \right) = \PP\left( {R = 1|Z=z,X} \right), &&  \pi \left( {X} \right) = \PP\left( {R = 1|X} \right), \\
    &\pi_0 = \PP\left( {R = 0} \right), && \rho_z(X) = \PP\left( {Z=z|X} \right), \\
    &{\mu _z}\left( X \right) = \EE\left\{ {Rh\left( {Y;{\psi _0}} \right)|Z=z,X} \right\}, && {\delta ^R}\left( X \right) = \pi_1(X) - \pi_0(X),\\
    &{\delta ^Y}\left( X \right) = \mu_1(X) - \mu_0(X), && \delta\left( X \right) = {\delta ^Y}\left( X \right) / {\delta ^R}\left( X \right).
\end{align*}
Note that $\delta(X)$ is a single-arm version of the Wald ratio estimand, a standard target of inference in instrumental variable causal inference literature, tailored here to address missing data.

To simplify the exposition, we first consider the binary IV setting in Section \ref{sec:idandest} and subsequently extend our results to the general IV setting in Section \ref{sec:generalIV}.

\section{Identification and Estimation for Binary IV}
\label{sec:idandest}

In this section, we present the identification and estimation results for the population mean when $Z$ is binary. These results essentially restate those of \cite{liu2025multiplicativeinstrumentalvariablemodel} and \cite{lee2025inferencenonlinearcounterfactualfunctionals} in missing-data notation; we refer the reader to their papers for detailed proofs.

\subsection{Identification of the Functional}
\label{subsec:id_binary}

Following \cite{liu2025multiplicativeinstrumentalvariablemodel} and \cite{lee2025inferencenonlinearcounterfactualfunctionals}, the missing population mean $\beta$ can be identified via the single-arm Wald ratio functional $\delta(X)$: 

    \[\beta  = {\EE_{X|R = 0}}\left[ \delta\left( X\right)|R = 0 \right].\]
    

Based on this identification result, one may in principle proceed to construct a plug-in estimator for $\beta$
    \[{{\hat \beta }_{ID}} = n_0^{ - 1}\sum\limits_{{R_i} = 0} {\hat\delta\left( X_i\right)}, \]
    where $n_0$ is the number of subjects with missing outcomes and ${\hat\delta}$ is an estimator of $\delta$.  
    However, even if ${\hat\delta}$ is consistent, using say flexible machine learning, the plug-in bias will in general be of first order with convergence rate substantially slower than $\sqrt{n}$ \citep{robins2017minimax}.  
    To address this limitation of the plug-in estimator, we develop a de-biased estimator which is guaranteed to have second-order bias, based on the influence function of $\beta$, using modern semiparametric theory \citep{robins2017minimax, chernozhukov2018double}. 

\subsection{Influence Function of the Functional}
\label{subsec:if_binary}
Following \cite{liu2025multiplicativeinstrumentalvariablemodel} and \cite{lee2025inferencenonlinearcounterfactualfunctionals}, the influence function (IF) of $\beta$ under a nonparametric model which places no restriction on the observed data distribution has the explicit form:

\begin{equation}
    \phi \left( {\beta ,O,P} \right) =\frac{2Z-1}{\rho_Z(X) }\cdot\frac{1 - \pi \left( {X} \right) }{\pi_0 \delta ^{R}\left(
    X\right)}\left[ Rh\left(
    Y;\psi _{0}\right) -R\delta \left( X\right) -\mu _{0}\left( X\right) +\pi
    _{0}\left( X\right) \delta \left( X\right) \right] +\frac{1-R}{\pi_0 }\left[ \delta \left( X\right) -\beta %
    \right]. \label{eq:IF_binary}
\end{equation}

Based on the derived IF, following \cite{liu2025multiplicativeinstrumentalvariablemodel} and \cite{lee2025inferencenonlinearcounterfactualfunctionals}, one may construct the semiparametric estimator for the mean functional among incomplete cases $\beta$.

    $$
    {{\hat \beta }_{IF}} = \frac{1}{n}\sum\limits_{i = 1}^n {\frac{{2{Z_i} - 1}}{{\hat \rho_{Z_i}\left(X_i \right)}}\cdot\frac{{1 - \hat\pi(X)}}{{\hat \pi_0}{{\hat \delta }^R}\left( {{X_i}} \right)}\left[ {Rh\left( {{Y_i}} \right) - R\hat \delta \left( {{X_i}} \right) - {{\hat \mu }_0}\left( {{X_i}} \right) + {{\hat \pi }_0}\left( {{X_i}} \right)\hat \delta \left( {{X_i}} \right)} \right]} + \frac{1}{n}\sum\limits_{i = 1}^n {\frac{{1 - {R_i}}}{{\hat \pi_0}}\cdot\hat \delta \left( {{X_i}} \right)} 
    $$
where $\hat\pi_z, \hat\pi \left( {X} \right), \hat\pi_0, \hat\rho_z(X), \hat{\mu _z}\left( X \right), {\hat\delta ^R}\left( X \right), {\hat\delta ^Y}\left( X \right), \hat\delta\left( X \right)$
are estimators of $\pi_z$, $\pi(X)$, $\pi_0$, $\rho_z(X)$, $\mu_z(X)$, 
$\delta^R(X)$, $\delta^Y(X)$, $\delta(X)$. Note that estimating the Wald-type ratio $\delta(X)$ can be unstable due to its ratio structure. 
To mitigate this issue, one may employ the Forster-Warmuth estimator 
for counterfactual regression problems of \cite{yang2023forster}. 
This estimator potentially attains the minimax rate of estimation of such a regression function under weaker conditions than a standard substitution-based estimator, see \cite{liu2025multiplicativeinstrumentalvariablemodel} and \cite{lee2025inferencenonlinearcounterfactualfunctionals} for details.


Proceeding as in \cite{liu2025multiplicativeinstrumentalvariablemodel} and \cite{lee2025inferencenonlinearcounterfactualfunctionals}, one can show that 
\begin{align}
\sqrt n \left( {\hat \beta  - \beta } \right) = &\underbrace {\sqrt n \int {\phi \left( {\beta ,O,P} \right)d{P_n}\left( O \right)} }_{{\rm{CLT}}} + \underbrace {\sqrt n R\left( {\hat P,P} \right)}_{{\rm{Second - order \, remainder}}} \nonumber \\
 &+ \underbrace {\sqrt n \int {\left\{ {\phi \left( {\beta ,O,\hat P} \right) - \phi \left( {\beta ,O,P} \right)} \right\}d\left\{ {{P_n}\left( O \right) - P\left( O \right)} \right\}} }_{{\rm{Empirical \, process}}}, \label{eq:decompostion}
\end{align}
where $P$ stands for the true distribution, $P_n$ is the empirical distribution from the observed data, $\hat{P}$ denotes the estimated distribution, and $\phi \left( {\beta,\cdot,\cdot} \right)$ is the influence function corresponding to $\hat{\beta}$. The first term takes the form of a standard sample average and is asymptotically normal by the Central Limit Theorem (CLT). The third term is an empirical process term, which is \( o_P(1) \) under certain Donsker conditions. However, this term can still be negligible through the use of cross-fitting and double machine learning techniques, even when the Donsker conditions are not satisfied, as we demonstrate later in Section~\ref{subsec:crossfitting}. The key component is the remainder term \( R(\hat{P}, P) \). The use of the IF guarantees that the corresponding estimator has a second-order remainder, which can be written as a sum of products of the estimation errors of nuisance functions. The following equation provides the explicit form of this remainder term, also characterized in \cite{liu2025multiplicativeinstrumentalvariablemodel} and \cite{lee2025inferencenonlinearcounterfactualfunctionals}.

\begin{align}
R\left( {\hat P,P} \right) = &{\EE_P}\left[ {\frac{1}{{{\pi _0}}}\left\{ {\hat \delta \left( X \right) - \delta \left( X \right)} \right\}\left\{ {\hat \pi \left( X \right) - \pi \left( X \right)} \right\}} \right] + {o_P}\left( {{n^{ - 1/2}}} \right) \nonumber\\
 &+ {\EE_P}\left[ {\frac{{1 - \hat \pi \left( X \right)}}{{{\pi _0}{{\hat \delta }^R}\left( X \right)}}\left\{ {\hat \delta \left( X \right) - \delta \left( X \right)} \right\}\left\{ {{{\hat \delta }^R}\left( X \right) - {\delta ^R}\left( X \right)} \right\}} \right] \nonumber\\
 &- {\EE_P}\left[ {\frac{{{\delta ^R}\left( X \right)\left\{ {1 - \hat \pi \left( X \right)} \right\}}}{{{\pi _0}{{\hat \delta }^R}\left( X \right){{\hat \rho }_1}\left( X \right)}}\left\{ {\hat \delta \left( X \right) - \delta \left( X \right)} \right\}\left\{ {{{\hat \rho }_0}\left( X \right) - {\rho _0}\left( X \right)} \right\}} \right] \nonumber\\
 &- {\EE_P}\left[ {\frac{{1 - \hat \pi \left( X \right)}}{{{\pi _0}{{\hat \delta }^R}\left( X \right){{\hat \rho }_1}\left( X \right){{\hat \rho }_0}\left( X \right)}}\left\{ {{{\hat \mu }_0}\left( X \right) - {\mu _0}\left( X \right)} \right\}\left\{ {{{\hat \rho }_0}\left( X \right) - {\rho _0}\left( X \right)} \right\}} \right] \nonumber\\
 &+ {\EE_P}\left[ {\frac{{\hat \delta \left( X \right)\left\{ {1 - \hat \pi \left( X \right)} \right\}}}{{{\pi _0}{{\hat \delta }^R}\left( X \right){{\hat \rho }_1}\left( X \right){{\hat \rho }_0}\left( X \right)}}\left\{ {{{\hat \pi }_0}\left( X \right) - {\pi _0}\left( X \right)} \right\}\left\{ {{{\hat \rho }_0}\left( X \right) - {\rho _0}\left( X \right)} \right\}} \right] \label{eq:rem_binary}
\end{align}




Therefore, 
if the 
    nuisance function estimators satisfy $R\left( {\hat P,P} \right) = {o_P}\left( {{n^{ - 1/2}}} \right)$, the IF-based estimator $\hat{\beta}_{IF}$ from sample splitting is asymptotically normal:
	\begin{align*}
		\sqrt{n}(\hat{\beta}_{IF}-\beta) \stackrel{D}{\longrightarrow} N(0,\sigma^2),
	\end{align*}
where $\sigma^2=\VV\left[\phi\left(\beta, O, P\right) \right]$. 





Following \cite{liu2025multiplicativeinstrumentalvariablemodel} and \cite{lee2025inferencenonlinearcounterfactualfunctionals}, the IF-based estimator has the multiple robustness property as another benefit. The IF $\phi \left( {\beta ,O,P} \right)$ is unbiased under the union of three models, i.e., if one of the following holds:
    \begin{enumerate}
        \item $\delta \left( X \right)$ is correct, and at least one of ${\mu _0}\left( X \right),{\pi _0}\left( X \right)$ and ${\mu _1}\left( X \right),{\pi _1}\left( X \right)$ is correct;
        \item ${\delta ^R}\left( X \right),\rho_Z(X),\pi(X)$ are correct;
        \item $\delta(X)$ and $\rho_Z(X)$ are correct.
    \end{enumerate}

\section{Extension to General Instrumental Variables}
\label{sec:generalIV}

In this section, we present new results that extend the multiplicative IV model for binary IV to a more general setting which accommodates more general forms of instruments, including polytomous, continuous, and multi-dimensional IVs, which are particularly common in missing data IV settings. Building on the approach from the previous section, we begin by presenting the identification formula for $\beta$ and then derive its IF. The IF derived for the binary IV setting can be shown to be recovered as a special case of the general formulation. We further construct the corresponding IF-based estimator and derive its remainder term, which we show is of second-order. The asymptotic properties of the estimator and its multiple robustness are also established based on the analysis of the remainder term.

For the general scenario, the following notations are used throughout this section:
\begin{align*}
    &\mu \left( X \right) = {\EE_{R,Y|X}}\left\{ {Rh\left( {Y;{\psi _0}} \right)|X} \right\}, &&  {\delta ^Y}\left( {Z,X} \right) = {\mu _Z}\left( X \right) - \mu \left( X \right), \\
    &{\delta ^R}\left( {Z,X} \right) = {\pi _Z}\left( X \right) - \pi \left( X \right), && {\delta }\left( {Z,X} \right) = {\delta^Y}\left( {Z,X} \right)/{\delta ^R}\left( {Z,X} \right).
\end{align*}

Note that the functions ${\delta ^Y}(Z,X)$, ${\delta ^R}(Z,X)$, and $\delta(Z,X)$ are defined as contrasts between quantities for a given level $Z$ and the corresponding marginal mean, rather than as differences between the two levels in the binary IV case. Consequently, they depend jointly on $Z$ and $X$, rather than on $X$ alone. While these general quantities cannot be directly reduced to the binary case, they are closely related. For example, in the proof of Corollary~\ref{coro:reduce} in the supplementary materials, we show that
\[
{\delta ^R}(Z = 1, X) = {\delta ^R}(X)\,\PP(Z = 0 \mid X).
\]

\subsection{Identification of the Functional}

The following theorem establishes that \emph{the generalized Wald-type functional} $\delta \left( {Z,X} \right)$ identifies the target parameter $\beta$.

\begin{theorem}
\label{thm:ID_general}
   The incomplete-case population mean parameter $\beta$ can be identified by 
    \[\beta  = {\EE_{Z,X|R = 0}}\left[ {\delta \left( {Z,X} \right)|R = 0} \right].\]
\end{theorem}

Analogous to Section \ref{subsec:id_binary}, a plug-in estimator based on the identification formula can be constructed as
\(
\hat{\beta}_{\mathrm{ID}} = n_0^{-1} \sum_{R_i = 0} \hat{\delta}(Z_i, X_i).
\)
However, analogous to the binary IV case, its bias will generally be of first order, motivating the following developments involving the corresponding IF which is guaranteed to have second order bias.

\subsection{Influence Function of the Functional}

The following result gives the IF for $\beta$ under the semiparametric model.

\begin{theorem}
\label{thm:IF_general}

    In the scenario of general IV $Z$, the influence function of the missing population mean $\beta$ is
\begin{align*}
\phi \left( {\beta ,O,P} \right) = \left\{ {g\left( {Z,X} \right) - g\left( X \right)} \right\}\left[ {Rh\left( Y \right) - \mu \left( {Z,X} \right) - \delta \left( {Z,X} \right)\left\{ {R - \pi \left( {Z,X} \right)} \right\}} \right] + \frac{{1 - R}}{{{\pi _0}}}\left\{ {\delta \left( {Z,X} \right) - \beta } \right\},
\end{align*}
where 
\begin{align*}
    &g\left( {Z,X} \right) = \frac{{1 - \pi \left( {Z,X} \right)}}{{{\pi _0}{\delta ^R}\left( {Z,X} \right)}}, \text{ and }  g\left( X \right) = {\EE_{Z|X}}\left[ {g\left( {Z,X} \right)|X} \right].
\end{align*}
\end{theorem}

It is worth noting that the IF in the previous theorem generalizes the result of binary IV, i.e., \eqref{eq:IF_binary} is a special case of Theorem \ref{thm:IF_general} when $Z$ is binary.

\begin{corollary}
\label{coro:reduce}
    The IF in Theorem \ref{thm:IF_general} is reduced to \eqref{eq:IF_binary} when $Z$ is binary.
\end{corollary}

Theorem \ref{thm:IF_general} provides the IF for the missing population mean $\beta$. 
Given this theorem, it is straightforward to derive the IF for the mean of the whole population, as stated in the following corollary.

\begin{corollary}
\label{coro:IFpopulation_general}
    In the scenario of general IV, the functional
    \[\EE\left[ {h\left( {Y;{\psi _0}} \right)} \right] = \PP\left( {R = 1} \right)\alpha  + \pi_0\beta \]
    as the population mean has IF given by 
    \begin{align*}
    &{\pi _0}\left\{ {g\left( {Z,X} \right) - g\left( X \right)} \right\}\left[ {Rh\left( Y \right) - \mu \left( {Z,X} \right) - \delta \left( {Z,X} \right)\left\{ {R - \pi \left( {Z,X} \right)} \right\}} \right]\\
     + &\alpha \left\{ {R - \PP\left( {R = 1} \right)} \right\} + \beta \left( {1 - R - {\pi _0}} \right) + R\left\{ {h\left( {Y;{\psi _0}} \right) - \alpha } \right\} + \left( {1 - R} \right)\left\{ {\delta \left( {Z,X} \right) - \beta } \right\}.
    \end{align*}   
\end{corollary}

Using IF, each functional can be semiparametrically estimated. For example, we estimate the missing population mean $\beta$ by

\[{{\hat \beta }_{IF}} = \frac{1}{n}\sum\limits_{i = 1}^n {\left\{ {\hat g\left( {{Z_i},{X_i}} \right) - \hat g\left( {{X_i}} \right)} \right\}\left[ {{R_i}h\left( {{Y_i}} \right) - \hat \mu \left( {{Z_i},{X_i}} \right) - \hat \delta \left( {{Z_i},{X_i}} \right)\left\{ {{R_i} - \hat \pi \left( {{Z_i},{X_i}} \right)} \right\}} \right] + \frac{{1 - {R_i}}}{{{{\hat \pi }_0}}} \cdot \hat \delta \left( {{Z_i},{X_i}} \right)}, \]
where $\hat g\left( {{Z_i},{X_i}} \right), \hat g\left( {{X_i}} \right), \hat \mu \left( {{Z_i},{X_i}} \right), \hat \delta \left( {{Z_i},{X_i}} \right), \hat \pi \left( {{Z_i},{X_i}} \right)$
are estimators of $g\left( {{Z_i},{X_i}} \right), g\left( {{X_i}} \right), \mu \left( {{Z_i},{X_i}} \right), \delta \left( {{Z_i},{X_i}} \right), \pi \left( {{Z_i},{X_i}} \right)$.

Next, we present another main contribution of our paper. The following theorem derives the explicit form of the remainder term for the IF-based estimator in the general IV setting and shows that it is of second order.

\begin{theorem}
\label{thm:rem_general}
    In the scenario of general IV, the IF-based estimator for the missing population mean ${{\hat \beta }_{IF}}$ has the second-order remainder term:
    \begin{align*}
R\left( {\hat P,P} \right) = &{\EE_P}\left[ {\frac{1}{{{\pi _0}}}\left\{ {\hat \delta \left( X \right) - \delta \left( X \right)} \right\}\left\{ {\hat \pi \left( {Z,X} \right) - \pi \left( {Z,X} \right)} \right\}} \right]\\
 &+ {\EE_P}\left[ {\hat g\left( {Z,X} \right)\left\{ {\hat \delta \left( X \right) - \delta \left( X \right)} \right\}\left\{ {{{\hat \delta }^R}\left( {Z,X} \right) - {\delta ^R}\left( {Z,X} \right)} \right\}} \right]\\
 &+ {\EE_P}\left[ {{\delta ^R}\left( {Z,X} \right)\left\{ {\hat \delta \left( X \right) - \delta \left( X \right)} \right\}\left[ {{\EE_{\hat P}}\left\{ {\hat g\left( {Z,X} \right)|X} \right\} - {\EE_P}\left\{ {\hat g\left( {Z,X} \right)|X} \right\}} \right]} \right]\\
 &+ {\EE_P}\left[ {\left\{ {\hat \mu \left( X \right) - \mu \left( X \right)} \right\}\left[ {{\EE_{\hat P}}\left\{ {\hat g\left( {Z,X} \right)|X} \right\} - {\EE_P}\left\{ {\hat g\left( {Z,X} \right)|X} \right\}} \right]} \right]\\
 &- {\EE_P}\left[ {\hat \delta \left( X \right)\left\{ {\hat \pi \left( X \right) - \pi \left( X \right)} \right\}\left[ {{\EE_{\hat P}}\left\{ {\hat g\left( {Z,X} \right)|X} \right\} - {\EE_P}\left\{ {\hat g\left( {Z,X} \right)|X} \right\}} \right]} \right].
    \end{align*}
\end{theorem}

The remainder term for the general IV case has a structure analogous to that of the binary IV case. Specifically, the quantity ${{\EE_{\hat P}}\left\{ {\hat g\left( {Z,X} \right)\mid X} \right\} - {\EE_P}\left\{ {\hat g\left( {Z,X} \right)\mid X} \right\}}$ in \eqref{eq:rem_binary} corresponds to ${{\hat \rho_0}(X) - \rho_0(X)}$ in Theorem~\ref{thm:rem_general}, both of which capture the error in estimating the conditional density of $Z$ given $X$.

The main advantage of having a second-order bias is that it requires much weaker conditions for the estimator's asymptotic normality. Specifically, to ensure that the remainder term is negligible, we need \( R(\hat{P}, P) = o_P(n^{-1/2}) \). For estimators not derived from the IF, such as the identification-based estimator, the remainder term is typically of the same order as the bias of the nuisance function estimators. Consequently, these nuisance functions must be estimated at a rate faster than \( \sqrt{n} \), which generally requires the use of parametric models. However, parametric models may be misspecified in practice, potentially introducing bias.

In contrast, nonparametric methods offer greater flexibility but usually cannot achieve \( \sqrt{n} \)-rate convergence. Fortunately, since the remainder term is second-order, it is sufficient for the product of the nuisance estimation rates to be faster than \( n^{-1/2} \), for instance, if each nuisance function is estimated at a rate faster than \( n^{-1/4} \). As a result, both the remainder and empirical process terms in the decomposition \eqref{eq:decompostion} become negligible, and only the sample average term remains, which is asymptotically normal by the CLT. The asymptotic properties of the IF-based estimator are formalized in the following corollary.


\begin{corollary}
\label{coro:rem_general}
    In the scenario of general IV, if the nuisance function estimators satisfy $R\left( {\hat P,P} \right) = {o_P}\left( {{n^{ - 1/2}}} \right)$, the IF-based estimator $\hat{\beta}_{IF}$ applying sample splitting in Section \ref{subsec:crossfitting} is asymptotically normal:
	\begin{align*}
		\sqrt{n}(\hat{\beta}_{IF}-\beta) \stackrel{D}{\longrightarrow} N(0,\sigma^2),
	\end{align*}
where $\sigma^2=\VV\left[\phi\left(\beta, O, P\right) \right]$
\end{corollary}

Additionally, the corollary provides a consistent variance estimator for the IF-based estimator
The variance can be estimated using the influence function, which enables valid inference procedures, including hypothesis testing and the construction of confidence intervals.

\begin{equation}
    \hat \VV\left( {{{\hat \beta }_{IF}}} \right) = {n^{ - 2}}\sum\limits_{i = 1}^n {\phi \left( {\beta ,{O_i},\hat P} \right)} . \label{eq:varest}
\end{equation}

Another benefit of the IF-based estimator is its multiple robustness, which also arises from the higher-order bias structure. When parametric models are used to estimate the nuisance functions, this property ensures that the estimator remains consistent even if some of the models are misspecified. The following corollary presents several scenarios under which the estimator is valid.

\begin{corollary}
\label{coro:multirobust_general}
    In the scenario of general $Z$, $\phi \left( {\beta ,O,P} \right)$ is unbiased under the union of three models, that is if one of the following holds:
    \begin{enumerate}
        \item $\delta \left( X \right),\mu \left( X \right),\pi \left( X \right)$ are correct;
        \item $\pi \left( {Z,X} \right),\rho_Z(X)$ are correct;
        \item $\delta(X)$ and $f(Z|X)$ are correct.
    \end{enumerate}
\end{corollary}

\subsection{Cross-Fitting and Double Machine Learning}
\label{subsec:crossfitting}

To derive estimators for the target functional, it is essential to estimate nuisance functions such as \( \mu(Z,X) \) and \( \pi(Z,X) \). Parametric methods are commonly used for this purpose; however, if some of the parametric models are misspecified, the resulting estimators will be inconsistent, as discussed in the previous section. To improve robustness and accuracy, nonparametric methods, including modern machine learning algorithms, have been applied to estimate these nuisance functions. Despite their flexibility, such methods may introduce regularization bias and overfitting, which can prevent the estimators from achieving the standard convergence rate. In particular, the empirical process term in \eqref{eq:decompostion} may fail to be \( o_P(1) \), causing the variance estimator in \eqref{eq:varest} to be invalid.

To address this issue, cross-fitting and double machine learning techniques have been developed to improve the robustness and accuracy of causal inference, particularly in econometrics and machine learning \citep{chernozhukov2018double}. Cross-fitting involves partitioning the data into multiple folds and using disjoint subsets for training and evaluation. This reduces overfitting and ensures that estimates are not overly dependent on a particular data subset. Double machine learning extends this idea by using machine learning algorithms to estimate nuisance functions in a two-step procedure: first, models are trained on the estimation fold to estimate the nuisance functions; second, these estimates are used to evaluate the influence function on the evaluation fold. This separation induces independence between the nuisance function estimation and the evaluation of the influence function, which helps ensure that the empirical process term in \eqref{eq:decompostion} becomes negligible. Algorithm~\ref{alg:cf} outlines the detailed steps for implementing the cross-fitted version of the IF-based estimator \( \hat{\beta}_{\mathrm{IF}} \).

\begin{algorithm}
\caption{Cross-Fitting to Derive \(\hat{\beta}_{IF}\)}\label{alg:cf}
\begin{algorithmic}
    \State \textbf{Input:} Dataset \(\{(Y_i, X_i, Z_i, R_i)\}_{i=1}^n\), number of folds \(K\).
    
    \State \textbf{Initialization:} Partition the data into \(K\) folds.
    
    \For{each fold \(k \in \{1, \ldots, K\}\)}
        \State Define training set \(\mathcal{I}_k^c\) (all data except the \(k\)-th fold) and validation set \(\mathcal{I}_k\) (the \(k\)-th fold).
        
        \State Estimate the nuisance functions using the training set \(\mathcal{I}_k^c\):
        $\hat g(Z_i, X_i),\allowbreak\ 
\hat g(X_i),\allowbreak\ 
\hat \mu(Z_i, X_i),\allowbreak\ 
\hat \delta(Z_i, X_i),\allowbreak\ 
\hat \pi(Z_i, X_i),\allowbreak\ 
\hat{\pi}_0$.
        
        \For{each \(i \in \mathcal{I}_k\)}
            \State Calculate the uncentered influence function component:
\begin{align*}
\tilde \phi \left( {\beta ,{O_i},\hat P} \right) = 
&\left\{ \hat g\left( Z_i, X_i \right) - \hat g\left( X_i \right) \right\}
\left[ R_i h\left( Y_i \right) - \hat \mu \left( Z_i, X_i \right) - \hat \delta \left( Z_i, X_i \right) \left\{ R_i - \hat \pi \left( Z_i, X_i \right) \right\} \right] \\
&\quad + \frac{1 - R_i}{\hat{\pi}_0} \cdot \hat \delta \left( Z_i, X_i \right)
\end{align*}

        \EndFor
    \EndFor  
    
    \State \textbf{Aggregate the results:}
    \State Calculate the mean of the influence function components over all data points:
    \[
    \hat{\beta}_{IF} = \frac{1}{n} \sum_{I=1}^n \tilde \phi \left( {\beta ,{O_i},\hat P} \right)
    \]
    
    \State \textbf{Output:} The cross-fitted estimator \(\hat{\beta}_{IF}\).
\end{algorithmic}
\end{algorithm}

\section{Simulation Study}
\label{sec:sim}

We conduct simulation studies to evaluate the performance of our proposed estimator. The simulation consists of two parts: the first considers a one-dimensional binary IV, while the second extends the setting to a general IV scenario with two binary IVs.

\subsection{Binary Instrumental Variable}

In the first part, we design a data-generating process involving a single binary IV. Two-dimensional covariates \( X = (X_1, X_2) \) are independently drawn from a uniform distribution \( U(0,1) \), and an unobserved confounder \( U \) is independently drawn from a normal distribution \( N(4, 0.5^2) \). The instrumental variable \( Z \) is generated from a logistic model conditional on \( X \). The missingness indicator \( R \) follows the proposed multiplicative IV model. The outcome \( Y \) is drawn from a normal distribution, with its expectation defined as a linear function of \( X \) modulated by a tilting term that depends on \( U \). The detailed data-generating mechanism is provided below.

\begin{align*}
P\left( {Z = 1|X} \right) &= \frac{{\exp \left( { - 1 + {X_1} + {X_2}} \right)}}{{1 + \exp \left( { - 1 + {X_1} + {X_2}} \right)}},\\
P\left( {R = 0|Z,U,X} \right) &= \exp \left\{ {\left( { - {X_1} - {X_2} - \frac{U}{4}} \right) + Z\left( {{X_1} + {X_2} + 1} \right)} \right\},\\
Y&\sim N\left( {\left( {{X_1} + {X_2}} \right) \cdot \exp \left( {\frac{U}{6}} \right),{{0.5}^2}} \right).
\end{align*}

We focus on estimating the mean of the missing outcome \( \beta \), whose true value is 1.8. The two estimators introduced in the paper—the identification-based estimator \( \hat\beta_{\mathrm{ID}} \) and the influence-function-based estimator \( \hat\beta_{\mathrm{IF}} \)—are used to estimate \( \beta \). 
For the IF-based estimators, we apply five-fold cross-fitting to reduce regularization bias and use a median adjustment over 11 repetitions to mitigate sensitivity to particular random splits. Variance of the IF-based estimators can be estimated via \eqref{eq:varest}, and the 95\% confidence interval can be derived correspondingly. 

We estimate the IV model using a generalized linear model with a logit link, incorporating all the covariates $X$. To estimate all remaining nuisance functions—including the missingness and outcome models—we employ the Super Learner algorithm, using both covariates and instrumental variables as inputs. The ensemble library consists of a variety of base learners: generalized linear models, multivariate adaptive regression splines, and multi-layer perceptrons. The sample sizes are set to 500, 750, and 1000, respectively, and each experiment is replicated 300 times. The degrees of freedom for the series of basis functions are set to 4, 5, and 6, respectively, depending on the sample size.

\begin{figure}[ht]
    \centering
    \includegraphics[width=0.8\textwidth]{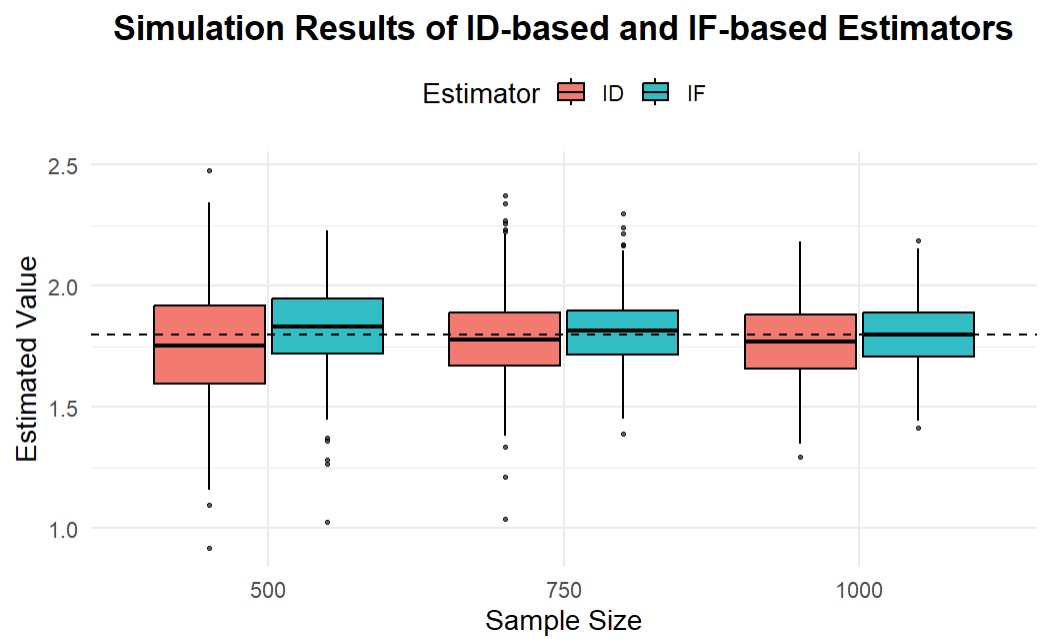}
    \caption{Boxplots of Estimators for the Binary Instrumental Variable Scenario Across Varying Sample Sizes. Dotted line indicates the true value.\label{fig:sim_binaryZ}}
\end{figure}

The results are shown in Figure \ref{fig:sim_binaryZ} and Table \ref{tab:sim_binaryZ}, which display boxplots and summarize the bias, variance, and mean squared error (MSE) of various estimators. Coverage rates for the 95\% confidence interval are reported only for the IF-based estimators, as a consistent variance estimator may not exist for the ID-based estimator. The recommended IF-based estimator generally exhibits smaller bias, variance, and MSE compared to the ID-based estimator, consistent with its theoretical asymptotic efficiency. 
The coverage rates of the IF-based estimators remain close to the nominal 95\%, supporting the consistency of the variance estimator derived from the influence function.

\begin{table}[ht]
\centering
\caption{Simulation Results for the Binary IV Scenario: Bias, Variance, and MSE of Estimators Across Different Sample Sizes. Coverage rates of IF-based estimators are also included.\label{tab:sim_binaryZ}}
\begin{tabular}{lcccccc}
\toprule
& \multicolumn{2}{c}{$n=500$} & \multicolumn{2}{c}{$n=750$} & \multicolumn{2}{c}{$n=1000$} \\
\cmidrule(lr){2-3} \cmidrule(lr){4-5} \cmidrule(lr){6-7}
& $\hat\beta_{ID}$ & $\hat\beta_{IF}$ 
& $\hat\beta_{ID}$ & $\hat\beta_{IF}$ 
& $\hat\beta_{ID}$ & $\hat\beta_{IF}$ \\
\midrule
Bias & -0.049 & 0.030 & -0.017 & 0.020 & -0.029 & -0.00064 \\
Var  & 0.058 & 0.029 & 0.036 & 0.020 & 0.030 & 0.018 \\
MSE  & 0.061 & 0.029 & 0.037 & 0.021 & 0.031 & 0.018 \\
Coverage Rate (\%) & / & 97.3 & / & 97.0 & / & 96.7 \\
\bottomrule
\end{tabular}
\end{table}


\subsection{General Instrumental Variable}

In the second part of our study, we consider a more general setting by introducing an additional binary IV into the data-generating process. The two-dimensional covariates \( X = (X_1, X_2) \) and the unmeasured confounder \( U \) are independently drawn from the standard uniform distribution \( U(0,1) \). The instrumental variables \( Z = (Z_1, Z_2) \) are both generated from logistic distributions. The observation indicator $R$ is still determined by a multiplicative IV model, now influenced by both \( Z_1 \) and \( Z_2 \). The outcome model for \( Y \) remains unchanged. The full data-generating formulas are provided below.

\begin{align*}
P\left( {{Z_1} = 1|X} \right) &= \frac{{\exp \left\{ {\left( { - 1 + {X_1} + {X_2}} \right)/4} \right\}}}{{1 + \exp \left\{ {\left( { - 1 + {X_1} + {X_2}} \right)/4} \right\}}}\\
P\left( {{Z_2} = 1|X} \right) &= \frac{{\exp \left\{ {\left( {{X_1} - {X_2}} \right)/4} \right\}}}{{1 + \exp \left\{ {\left( {{X_1} - {X_2}} \right)/4} \right\}}}\\
P\left( {R = 0|Z,U,X} \right) &= \exp \left[ {\frac{1}{4}\left\{ {8 + {X_1} - {X_2} - U + {Z_1}\left( { - 1 - {X_1} - {X_2}} \right) + {Z_2}\left( {8 + {X_1} - {X_2}} \right)} \right\}} \right]
\end{align*}

The target estimand is again the mean of the missing outcome \( \beta \), with the true value now set to 1.07. 
The two estimators \( \hat\beta_{\mathrm{ID}} \) and \( \hat\beta_{\mathrm{IF}} \) are implemented in the same manner, using the same basis functions in the Super Learner algorithm and identical tuning parameters. Since there are \( 2 \times 2 = 4 \) possible levels of the two binary IVs, additional nuisance functions are required for prediction. The sample size is increased to 10,000.

\begin{figure}[ht]
    \centering
    \includegraphics[width=0.8\textwidth]{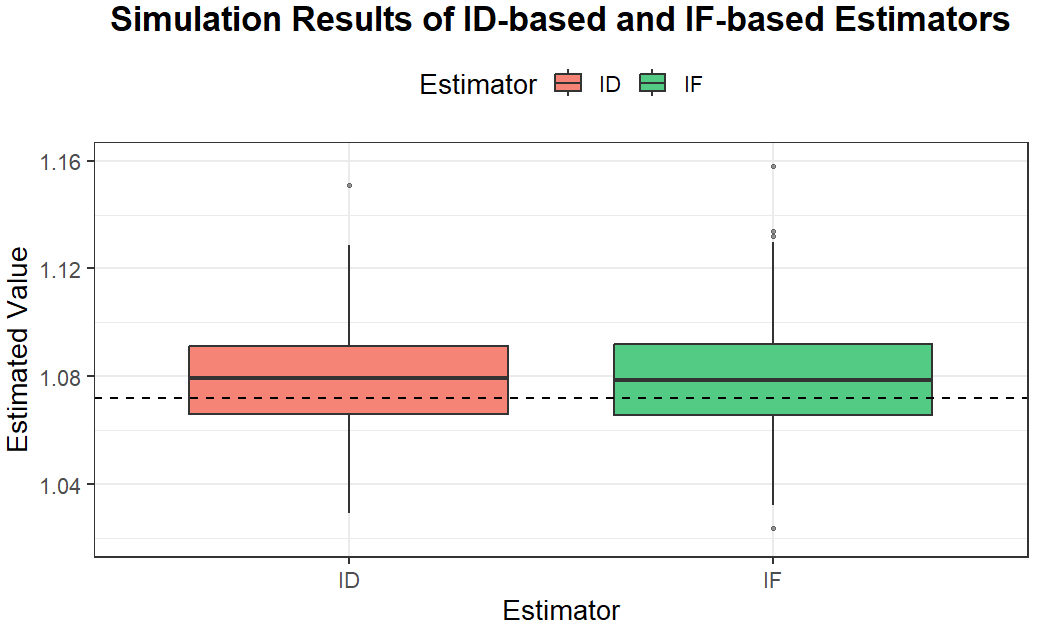}
    \caption{Boxplots of Estimators in the General Instrumental Variable Scenario. The dashed line indicates the true value. 
    \label{fig:sim_generalZ}}
\end{figure}

Figure~\ref{fig:sim_generalZ} and Table~\ref{tab:sim_generalZ} present the simulation results for the general IV scenario. The IF-based estimator exhibits slightly smaller bias but slightly larger variance. 
Compared to the binary IV scenario, the differences among the two estimators are less pronounced. One possible explanation for the limited advantage of the IF-based estimators over the ID-based estimator is that the general IV setting involves more levels, making the numerator \( \delta^R(X,Z) \) more likely to approach zero. This increases numerical instability may also lead to an overestimation of the variance by the IF-based method. A potential remedy is to remove extreme outliers in the IF values across the dataset.

\begin{table}[ht]
\centering
\caption{Bias, Variance, MSE, and Coverage Rate of Estimators in the General Instrumental Variable Scenario.\label{tab:sim_generalZ}}
\begin{tabular}{lcc}
\toprule
& $\hat\beta_{ID}$ & $\hat\beta_{IF}$ \\
\midrule
Bias           & 0.00718  & 0.00688  \\
Variance       & 0.000350 & 0.000403 \\
MSE            & 0.000400 & 0.000449 \\
Coverage Rate (\%) & /        & 96.7     \\
\bottomrule
\end{tabular}
\end{table}


\section{Real Data Application}
\label{sec:realdat}

To illustrate the proposed IV method, we examine data from a household survey conducted in Mochudi, Botswana, designed to estimate HIV seroprevalence among adults while addressing selective nonresponse in HIV testing. The survey included 4,997 adults aged 16 to 64, of whom 4,045 (81\%) provided complete HIV test results. Among the remaining 952 individuals with missing HIV status ($R = 0$), 111 (2\%) consented to testing but lacked final results, and 841 (17\%) explicitly refused to participate in the HIV testing component. Such nonparticipation, even among those successfully contacted, presents a likely source of selection bias.

The dataset contains fully observed covariates, including participant gender, age, and the number of nights spent away from home in the past month, which is an established risk factor for HIV. We consider interviewer characteristics as candidate instruments: gender, age, and years of experience. These attributes are expected to influence an individual's likelihood of completing HIV testing, but are unlikely to directly affect HIV status, especially given that interviewer assignments were randomized prior to the survey. 

Potential unmeasured confounders, such as the health-seeking behavior of the participants, may jointly affect both their HIV status and their likelihood of responding to the test. Since these confounders pertain to participants rather than interviewers, it is reasonable to assume that they do not interact with interviewer-level instruments in the response mechanism. This supports the assumption of a separable and independent response model, making the multiplicative IV framework appropriate for this setting.

We estimate HIV seroprevalence using our proposed estimator based on the influence function. Since the instrumental variable is a three-dimensional vector, we apply the generalized IV approach described in Section \ref{sec:generalIV}. To facilitate analysis, the two continuous instruments—age and years of experience—are discretized into categorical variables based on their quartiles.

We estimate the IV model using a generalized linear model with a logit link, incorporating the relevant covariates. To estimate the remaining nuisance functions from the missingness and outcome models, we employ the Super Learner algorithm, using both covariates and instrumental variables as inputs. The ensemble library includes a range of basis learners: generalized linear models, multivariate adaptive regression splines, generalized additive models, polynomial splines, random forests, single-layer perceptrons, and multi-layer perceptrons. To mitigate overfitting and reduce bias from machine learning-based estimation, we apply five-fold cross-fitting throughout the procedure.

Our estimate of HIV seroprevalence among nonrespondents is 38.4\% (95\% CI: 33.1\%--43.7\%), substantially higher than the observed prevalence among respondents, which is 21.4\% (95\% CI: 20.2\%--22.7\%). This discrepancy highlights a significant degree of selection bias when nonrespondents are ignored. The estimated HIV seroprevalence for the overall adult population in Mochudi is 24.7\% (95\% CI: 18.6\%--30.8\%), which aligns closely with findings from previous studies, such as the 25.8\% estimate reported by Sun et al. (2018).

\section{Discussion}
\label{sec:discussion}

In this paper, we proposed a novel approach to handle nonignorable missing data by introducing a multiplicative selection model that incorporates instrumental variables and allows for the presence of unmeasured confounders. By leveraging semiparametric theory, we derived the influence function and constructed an estimator with desirable properties such as multiple robustness, and interpretability. Our framework generalizes existing work by extending from binary to general instrumental variables, including multi-level, continuous, and multi-dimensional cases. Simulation studies and a real-world application to an HIV seroprevalence survey in Botswana demonstrate the practical utility of our method. 

It is important to note that the IF we derived for the general IV setting is inefficient, unlike in the binary IV case. The inefficiency arises because the target parameter is over-identified when the IV is polytomous, continuous, or multi-dimensional. Consequently, multiple identification formulas lead to multiple candidate IFs, and the one we obtained is only a particular instance that may not coincide with the efficient influence function (EIF). Obtaining the EIF requires projecting the IF onto the tangent space of the multiplicative IV model. In the supplementary material, we characterized this tangent space under the imposed restrictions. However, this derivation is highly non-trivial and structurally complex, which makes the projection extremely difficult and the resulting efficient estimator impractical for use. Nevertheless, although our estimator does not attain the semiparametric efficiency bound, it still provides the important advantage of reducing higher-order asymptotic bias.

Future research directions include extending the proposed model to longitudinal and developing tools for sensitivity analysis under possible violations of the model assumptions. Although the proposed estimator has desirable asymptotic properties, its finite-sample performance may vary. In particular, unstable outliers can occur when the denominators in the ratio structure of the Wald-type statistics are close to zero, which is especially common in the general IV setting where ratios must be evaluated for each IV level. Studying the finite-sample behavior of the estimator and identifying potential remedies to improve its stability would be an interesting direction for future work.

\bibliographystyle{plainnat}
\bibliography{ci}

\begin{thebibliography}{28}
\providecommand{\natexlab}[1]{#1}
\providecommand{\url}[1]{\texttt{#1}}
\expandafter\ifx\csname urlstyle\endcsname\relax
  \providecommand{\doi}[1]{doi: #1}\else
  \providecommand{\doi}{doi: \begingroup \urlstyle{rm}\Url}\fi

\bibitem[Chernozhukov et~al.(2018)Chernozhukov, Chetverikov, Demirer, Duflo, Hansen, Newey, and Robins]{chernozhukov2018double}
Victor Chernozhukov, Denis Chetverikov, Mert Demirer, Esther Duflo, Christian Hansen, Whitney Newey, and James Robins.
\newblock Double/debiased machine learning for treatment and structural parameters.
\newblock \emph{The Econometrics Journal}, 21:\penalty0 1--68, 2018.

\bibitem[Das et~al.(2003)Das, Newey, and Vella]{das2003nonparametric}
Mitali Das, Whitney~K Newey, and Francis Vella.
\newblock Nonparametric estimation of sample selection models.
\newblock \emph{The Review of Economic Studies}, 70:\penalty0 33--58, 2003.

\bibitem[Diggle and Kenward(1994)]{diggle1994informative}
Peter Diggle and Michael~G Kenward.
\newblock Informative drop-out in longitudinal data analysis.
\newblock \emph{Journal of the Royal Statistical Society Series C: Applied Statistics}, 43\penalty0 (1):\penalty0 49--73, 1994.

\bibitem[Heckman(1997)]{heckman1997instrumental}
James Heckman.
\newblock Instrumental variables: A study of implicit behavioral assumptions used in making program evaluations.
\newblock \emph{J Hum Resour}, 32:\penalty0 441--462, 1997.

\bibitem[Heckman(1979)]{heckman1979sample}
James~J Heckman.
\newblock Sample selection bias as a specification error.
\newblock \emph{Econometrica}, 47:\penalty0 153--161, 1979.

\bibitem[Kott(2014)]{kott2014calibration}
Phillip~S Kott.
\newblock Calibration weighting when model and calibration variables can differ.
\newblock In L.~P. Conti \& G. M.~Ranalli F.~Mecatti, editor, \emph{Contributions to Sampling Statistics}, pages 1--18. Cham: Springer, 2014.

\bibitem[Lee et~al.(2025)Lee, Yu, Liu, Park, Zhang, Robins, and Tchetgen]{lee2025inferencenonlinearcounterfactualfunctionals}
Yonghoon Lee, Mengxin Yu, Jiewen Liu, Chan Park, Yunshu Zhang, James~M. Robins, and Eric J.~Tchetgen Tchetgen.
\newblock Inference on nonlinear counterfactual functionals under a multiplicative iv model, 2025.
\newblock URL \url{https://arxiv.org/abs/2507.15612}.

\bibitem[Li et~al.(2023)Li, Miao, and Tchetgen~Tchetgen]{li2023non}
Wei Li, Wang Miao, and Eric Tchetgen~Tchetgen.
\newblock Non-parametric inference about mean functionals of non-ignorable non-response data without identifying the joint distribution.
\newblock \emph{Journal of the Royal Statistical Society Series B: Statistical Methodology}, 85\penalty0 (3):\penalty0 913--935, 2023.

\bibitem[Liu et~al.(2025)Liu, Park, Lee, Zhang, Yu, Robins, and Tchetgen]{liu2025multiplicativeinstrumentalvariablemodel}
Jiewen Liu, Chan Park, Yonghoon Lee, Yunshu Zhang, Mengxin Yu, James~M. Robins, and Eric J.~Tchetgen Tchetgen.
\newblock The multiplicative instrumental variable model, 2025.
\newblock URL \url{https://arxiv.org/abs/2507.09302}.

\bibitem[Miao and Tchetgen~Tchetgen(2016)]{miao2016varieties}
Wang Miao and E.~J. Tchetgen~Tchetgen.
\newblock On varieties of doubly robust estimators under missingness not at random with a shadow variable.
\newblock \emph{Biometrika}, 2:\penalty0 475--482, 2016.

\bibitem[Miao et~al.(2016)Miao, Ding, and Geng]{miao2016identifiability}
Wang Miao, Peng Ding, and Zhi Geng.
\newblock Identifiability of normal and normal mixture models with nonignorable missing data.
\newblock \emph{J Am Stat Assoc}, 111:\penalty0 1673--1683, 2016.

\bibitem[Miao et~al.(2024)Miao, Liu, Li, Tchetgen~Tchetgen, and Geng]{miao2024identification}
Wang Miao, Lan Liu, Yilin Li, Eric~J Tchetgen~Tchetgen, and Zhi Geng.
\newblock Identification and semiparametric efficiency theory of nonignorable missing data with a shadow variable.
\newblock \emph{ACM/JMS Journal of Data Science}, 1\penalty0 (2):\penalty0 1--23, 2024.

\bibitem[Newey et~al.(1990)Newey, Powell, and Walker]{newey1990semiparametricb}
Whitney~K Newey, James~L Powell, and James~R Walker.
\newblock Semiparametric estimation of selection models: some empirical results.
\newblock \emph{The american economic review}, 80\penalty0 (2):\penalty0 324--328, 1990.

\bibitem[Novitsky et~al.(2015)Novitsky, K{\"u}hnert, Moyo, Widenfelt, Okui, and Essex]{novitsky2015phylodynamic}
Vlad Novitsky, Denise K{\"u}hnert, Sikhulile Moyo, Erik Widenfelt, Lillian Okui, and Max Essex.
\newblock Phylodynamic analysis of hiv sub-epidemics in mochudi, botswana.
\newblock \emph{Epidemics}, 13:\penalty0 44--55, 2015.

\bibitem[Robins et~al.(1994)Robins, Rotnitzky, and Zhao]{robins1994estimation}
James~M Robins, Andrea Rotnitzky, and Lue~Ping Zhao.
\newblock Estimation of regression coefficients when some regressors are not always observed.
\newblock \emph{Journal of the American statistical Association}, 89\penalty0 (427):\penalty0 846--866, 1994.

\bibitem[Robins et~al.(2000)Robins, Rotnitzky, and Scharfstein]{robins2000sensitivity}
James~M Robins, Andrea Rotnitzky, and Daniel~O Scharfstein.
\newblock Sensitivity analysis for selection bias and unmeasured confounding in missing data and causal inference models.
\newblock In \emph{Statistical Models in Epidemiology, the Environment, and Clinical Trials}, pages 1--94. Springer, New York: Springer, 2000.

\bibitem[Robins et~al.(2017)Robins, Li, Tchetgen~Tchetgen, Mukherjee, and van~der Vaart]{robins2017minimax}
James~M. Robins, Lingling Li, Eric Tchetgen~Tchetgen, Rajarshi Mukherjee, and Aad~W. van~der Vaart.
\newblock Minimax estimation of a functional on a structured high-dimensional model.
\newblock \emph{Annals of Statistics}, 45\penalty0 (5):\penalty0 1951--1987, 2017.
\newblock \doi{10.1214/16-AOS1515}.

\bibitem[Rotnitzky and Robins(1997)]{rotnitzky1997analysis}
Andrea Rotnitzky and James Robins.
\newblock Analysis of semi-parametric regression models with non-ignorable non-response.
\newblock \emph{Statistics in medicine}, 16\penalty0 (1):\penalty0 81--102, 1997.

\bibitem[Roy(2003)]{roy2003modeling}
Jason Roy.
\newblock Modeling longitudinal data with nonignorable dropouts using a latent dropout class model.
\newblock \emph{Biometrics}, 59\penalty0 (4):\penalty0 829--836, 2003.

\bibitem[Rubin(1976)]{rubin1976inference}
Donald~B Rubin.
\newblock Inference and missing data.
\newblock \emph{Biometrika}, 63:\penalty0 581--592, 1976.

\bibitem[Rubin(2018)]{rubin2018multiple}
Donald~B Rubin.
\newblock Multiple imputation.
\newblock In \emph{Flexible imputation of missing data, second edition}, pages 29--62. Chapman and Hall/CRC, 2018.

\bibitem[Seaman et~al.(2012)Seaman, White, Copas, and Li]{seaman2012combining}
Shaun~R Seaman, Ian~R White, Andrew~J Copas, and Leah Li.
\newblock Combining multiple imputation and inverse-probability weighting.
\newblock \emph{Biometrics}, 68\penalty0 (1):\penalty0 129--137, 2012.

\bibitem[Sun et~al.(2018)Sun, Liu, Miao, Wirth, Robins, and Tchetgen]{sun2018semiparametric}
BaoLuo Sun, Lan Liu, Wang Miao, Kathleen Wirth, James Robins, and Eric J~Tchetgen Tchetgen.
\newblock Semiparametric estimation with data missing not at random using an instrumental variable.
\newblock \emph{Statistica Sinica}, 28\penalty0 (4):\penalty0 1965, 2018.

\bibitem[Tchetgen~Tchetgen and Wirth(2017)]{tchetgen2017general}
Eric~J Tchetgen~Tchetgen and Kathleen~E Wirth.
\newblock A general instrumental variable framework for regression analysis with outcome missing not at random.
\newblock \emph{Biometrics}, 73:\penalty0 1123--1131, 2017.

\bibitem[Tsiatis(2006)]{tsiatis2006semiparametric}
Anastasios~A Tsiatis.
\newblock \emph{Semiparametric theory and missing data}.
\newblock Springer, 2006.

\bibitem[Van~der Laan et~al.(2011)Van~der Laan, Rose, et~al.]{van2011targeted}
Mark~J Van~der Laan, Sherri Rose, et~al.
\newblock \emph{Targeted learning: causal inference for observational and experimental data}, volume~4.
\newblock Springer, 2011.

\bibitem[Wu and Carroll(1988)]{wu1988estimation}
Margaret~C Wu and Raymond~J Carroll.
\newblock Estimation and comparison of changes in the presence of informative right censoring by modeling the censoring process.
\newblock \emph{Biometrics}, pages 175--188, 1988.

\bibitem[Yang et~al.(2023)Yang, Kuchibhotla, and Tchetgen]{yang2023forster}
Yachong Yang, Arun~Kumar Kuchibhotla, and Eric~Tchetgen Tchetgen.
\newblock Forster-warmuth counterfactual regression: A unified learning approach.
\newblock \emph{arXiv preprint arXiv:2307.16798}, 2023.

\end{thebibliography}

\end{document}